\input{aipcheck}

\documentclass[
    ,final            
  ]
  {aipproc}

\layoutstyle{8x11double}

\newcommand{\be}{\begin{equation}}
\newcommand{\ee}{\end{equation}}

\newcommand{\eq}[1]{Eq. (\ref{#1})} 
 
\newcommand{\Fig}[1]{Fig. (\ref{#1})} 
 \newcommand{\eqa}{\begin{eqnarray}}
\newcommand{\eeq}{\end{eqnarray}}

\begin{document}

\title{3D Hall MHD Modeling of Solar Wind Plasma Spectra}



\author{Dastgeer Shaikh\footnote{\tt Email:dastgeer.shaikh@uah.edu}}{
  address={Department of Physics and Center for Space Plasma and Aeronomy Research (CSPAR), \\
The University of Alabama in Huntsville, Huntsville, AL 35899, USA}
}

\author{G. P. Zank}{}

\begin{abstract}                                         
We present fully self consistent 3D simulations of compressible Hall
MHD plasma that describe spectral features relevant to the solar wind
plasma. We find that a $k^{-7/3}$ spectrum sets in for the
fluctuations that are smaller than ion gyro radius.  We further
investigate scale dependent anisotropy led by nonlinear processes
relevant to the solar wind plasma. Our work is important particularly
in understanding the role of wave and nonlinear cascades in the
evolution of the solar wind, structure formation at the largest
scales. 
\end{abstract}

\maketitle


\section{1. introduction}
Solar wind (SW) fluctuations comprise a multitude of length and time
scales that collectively exhibit numerous nonlinear processes. Despite
it's complex evolutionary dynamics, solar wind provides the best
laboratory for testing many nonlinear theories and simulation
models. Spacecraft databases provide compelling observational
evidences that the inertial range magnetic field fluctuations
associated with characteristic frequencies, that are smaller than the
ion gyro frequency ($\Omega_{ci}$), can be described predominantly by
a Kolmogorov-like 5/3 spectrum \cite{Goldstein1995}. Theoretic
analysis describing the 5/3 SW spectrum regime relies largely on the
usual magnetohydrodynamic (MHD) model of plasma.  Intriguingly, the
higher time resolution databases of solar wind fluctuations depict a
spectral break near the end of the 5/3 spectrum that corresponds to a
high frequency ($>\Omega_{ci}$) regime where turbulent cascades are
{\em not} explainable by the usual MHD models. This refers to a second
inertial range where turbulent cascades follow a
$k^{-7/3}$ \cite{leamon} (where $k$ is a wavenumber) spectrum in which
the characteristic fluctuations evolve typically on kinetic Alfven
time scales. The onset of the second or the kinetic Alfven inertial
range is still elusive to our understanding of SW turbulence and the
issue has been under a constant debate since many years. The mechanism
leading to the spectral break has been thought to be either mediated
by the kinetic Alfven waves (KAWs) \cite{hasegawa}, or damping of ion
cyclotron waves, or dispersive processes \cite{gary}, Hall
effects \cite{alex,shaikh09}.

In this paper, we describe results from our three dimensional
simulations that explains that the $k^{-7/3}$ spectrum observed above
the spectral break in SW may be led by the Hall effects in the KAW
regime.  The underlying model, in our simulation model, is based on a
two fluid Hall MHD description of plasma that consists of both
electrons and ions. In the high frequency regime, $\omega
> \Omega_{ci}$, the inertialess electrons contribute to the electric
field which is dominated essentially by the Hall term corresponding to
${\bf J}
\times {\bf B}$ force.  The latter, upon substituting in the ion
momentum equation, modifies ion momentum, magnetic field and total
energy in a manner to introduce a high frequency ($\omega >
\Omega_{ci}$) and small scale ($k_\perp \rho_L>1$, where $\rho_L$ is
ion Larmour radius) plasma motion. The characteristic length scales
($k^{-1}$) associated with the plasma motions are smaller than ion
gyro radii ($\rho_L$).  The quasi-neutral solar wind plasma density
($\rho$), velocity (${\bf U}$), magnetic field (${\bf B}$) and total
pressure ($P=P_e+P_i$) fluctuations can then be cast into a set of
Hall MHD equations as follows (see Ref \cite{shaikh09} for detail description).
\be
\label{mhd:cont}
\frac{\partial \rho}{\partial t} + \nabla \cdot (\rho{\bf U})=0,
\ee
\be
\label{mhd:mom}
\rho \left(  \frac{\partial }{\partial t} + {\bf U} \cdot \nabla \right) {\bf U}
= -\nabla P + \frac{1}{c} {\bf J} \times {\bf B}+ 
\mu \left(\nabla^2 {\bf U} + \frac{1}{3}\nabla{\nabla \cdot \bf U}\right)
\ee
\be
\label{mhd:mag}
 \frac{\partial {\bf B}}{\partial t} = \nabla \times \left({\bf U} \times {\bf B}-
d_i \frac{{\bf J} \times {\bf B}}{\rho} \right)+\eta \nabla^2 {\bf B},
\ee
\be
\label{mhd:en}
 \frac{\partial e}{\partial t} + \nabla \cdot \left( \frac{1}{2}\rho
 U^2{\bf U} + \frac{\gamma P}{\gamma-1}{\bf U} +
 \frac{c}{4\pi}{\bf E} \times {\bf B} \right) =0 \ee 
where $e=1/2\rho U^2 + P/(\gamma-1)+B^2/8\pi$ is a total energy of
plasma that contains both electron and ion motions.  All the dynamical
variables are functions of three space and a time, i.e. $(x,y,z,t)$,
co-ordinates. Equations (\ref{mhd:cont}) to (\ref{mhd:en}) are
normalized by typical length $\ell_0$ and time $t_0 = \ell_0/v_0$
scales in our simulations, where $v_0=B_0/(4\pi \rho_0)^{1/2}$ is
Alfv\'en velocity such that $\bar{\nabla}=\ell_0{\nabla},
\partial/\partial \bar{t}=t_0\partial/\partial t, \bar{\bf U}={\bf
  U}/v_0,\bar{\bf B}={\bf B}/v_0(4\pi \rho_0)^{1/2},
\bar{P}=P/\rho_0v_0^2, \bar{\rho}=\rho/\rho_0$.  The parameters $\mu$
and $\eta$ correspond respectively to ion-electron viscous drag term
and magnetic field diffusivity. While the viscous drag modifies the
dissipation in plasma momentum in a nonlinear manner, the magnetic
diffusion damps the small scale magnetic field fluctuations linearly.
The magnetic field is measured in the unit of Alfv\'en velocity. The
dimensionless parameter in magnetic field \eq{mhd:mag} i.e. ion skin
depth $\bar{d}_i=d_i/\ell_0, d_i=C/\omega_{pi}$ is associated with the
Hall term.  This means the ion inertial scale length ($d_i$) is a
natural or an intrinsic length scale present in the Hall MHD model
which accounts for finite Larmour radius effects corresponding to high
frequency oscillations in $kd_i>1$ regime.  Clearly Hall forces
dominate the magnetoplasma dynamics when $1/\rho({\bf J} \times {\bf
  B}) > {\bf U} \times {\bf B}$ term in \eq{mhd:mag} which
in turn introduces time scales corresponding to the high frequency
plasma fluctuations in $kd_i>1$ regime.  Furthermore, our model
includes a full energy equation [\eq{mhd:en}] unlike an adiabatic
relation between the pressure and density. The use of energy equation
enables us to study a self-consistent evolution of turbulent heating
resulting from nonlinear energy cascades in the solar wind plasma.

\begin{figure}[t]
\label{spectra}
  \includegraphics[height=.25\textheight]{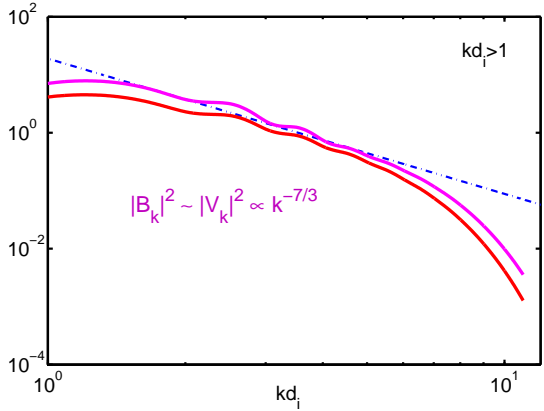} \caption{Inertial
  range turbulent spectra for magnetic and velocity field
  fluctuations. The fluctuations closely follow respectively
  $k^{-7/3}$ scalings in the $kd_i>1$ KAW regime.}
\end{figure}

\section{2. Turbulent  spectra}
To study the nonlinear evolution of turbulent cascades in a Hall MHD
solar wind plasma, we have developed a fully three dimensional
compressible Hall MHD code. Our code is massively parallelized using
Message Passing Interface (MPI) to run on cluster like
distributed-memory supercomputers. The code is scalable and
transportable on different cluster machines. The spatial
descritization employs a pseudospectral algorithm \cite{scheme} based
on a Fourier harmonic expansion of the bases for physical variables
(i.e. the density, magnetic field, velocity, temperature and energy)
whereas the temporal integration uses a Runge Kutta (RK) 4th order
method.  The boundary conditions are periodic along the $x,y$ and $z$
directions in the local rectangular region of the solar wind plasma.

\begin{figure}[t]
\label{aniso}
  \includegraphics[height=.25\textheight]{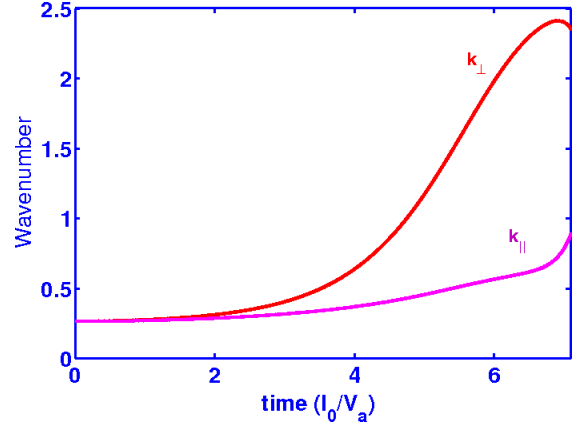}
  \caption{Evolution of $\langle k_\perp(t)\rangle $ and $
  \langle k_\parallel(t) \rangle $ as a measure of anisotropy in Hall
  MHD. Initially, $ \langle k_\perp(t=0) \rangle =
  k_\parallel(t=0)$. As time evolves, anisotropy in spectral transfer
  progressively develops such that $\langle k_\perp(t)\rangle >
  \langle k_\parallel(t) \rangle $.}
\end{figure}

The turbulent fluctuations are initialized by using a uniform
isotropic random spectral distribution of Fourier modes concentrated
in a smaller band of lower wavenumbers ($k<0.1~k_{max}$). While
spectral amplitude of the fluctuations is random for each Fourier
coefficient, it follows a certain initial spectral distribution
proportional to $k^{-\alpha}$, where $\alpha$ is initial spectral
index.  The spectral distribution set up in this manner initializes
random scale turbulent fluctuations.  A constant background magnetic
field is included along the $z$ direction to deal primarily with the
large scale or background solar wind magnetic field.  Turbulent
fluctuations in our simulations are driven either at the lowest
Fourier modes or evolve freely under the influence of self-consistent
dynamics described by the set of Eqs. (\ref{mhd:cont}) to
(\ref{mhd:en}). The inertial range spectral cascades in the either
cases lead to the nearly identical turbulent spectra. We have further
carried out simulations for a range of various parameters and spectral
distributions to ensure the validity of our codes and the physical
results. The simulation parameters are; spectral resolution is
$128^3$, $\eta=\mu=10^{-3}, \beta=1.0, kd_i \sim 0.1-10,
L_x=L_y=L_z=2\pi$.  The nonlinear coupling of velocity and magnetic
field fluctuations, amidst density perturbations, excites
high-frequency and short wavelength (by the $\omega/\omega_{ci}$
effect) compressional dispersive KAWs. The nonlinear spectral cascade
in the modified KAW regime leads to a secondary inertial range in the
vicinity of $kd_i \simeq 1$, where the turbulent magnetic and velocity
fluctuations form spectra close to
$k^{-7/3}$ \cite{iros,kol,krai}. This is displayed
in \Fig{spectra}. It is shown in Shaikh \& Zank (2009) that the
spectra described in Fig (1) is led by Hall effects.

\begin{figure}[t]
\label{aniso2}
  \includegraphics[height=.27\textheight]{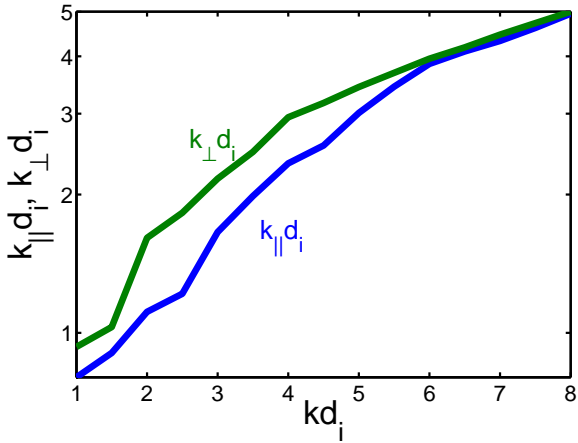}
  \caption{ Spectrum of $ k_\perp$ and $ k_\parallel$ as a
  function of $k$.  The large scale inertial range turbulent
  fluctuations are more anisotropic as compared to the smaller ones.}
\end{figure}

\begin{figure}[t]
\label{angle}
  \includegraphics[height=.27\textheight]{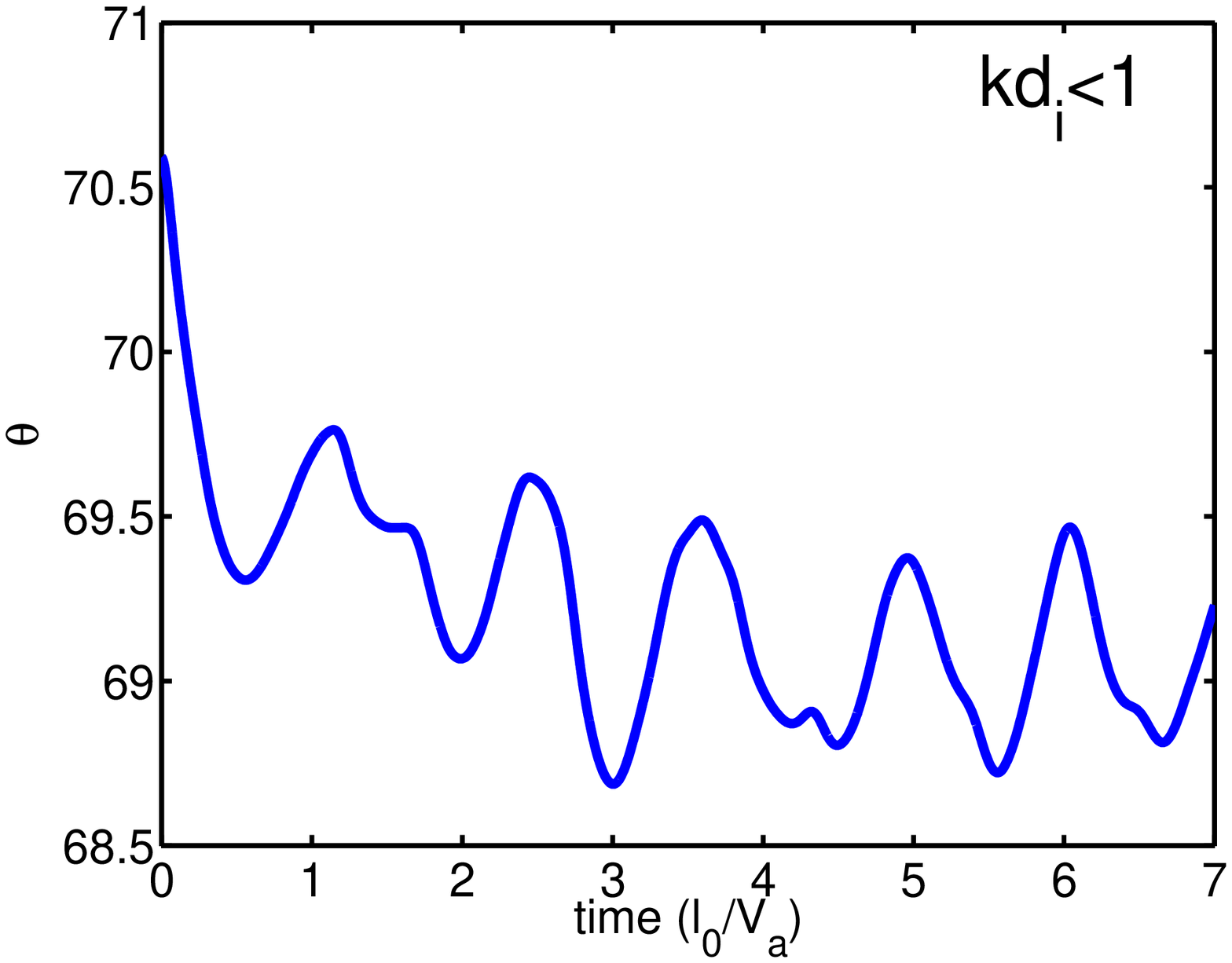}
\includegraphics[height=.26\textheight]{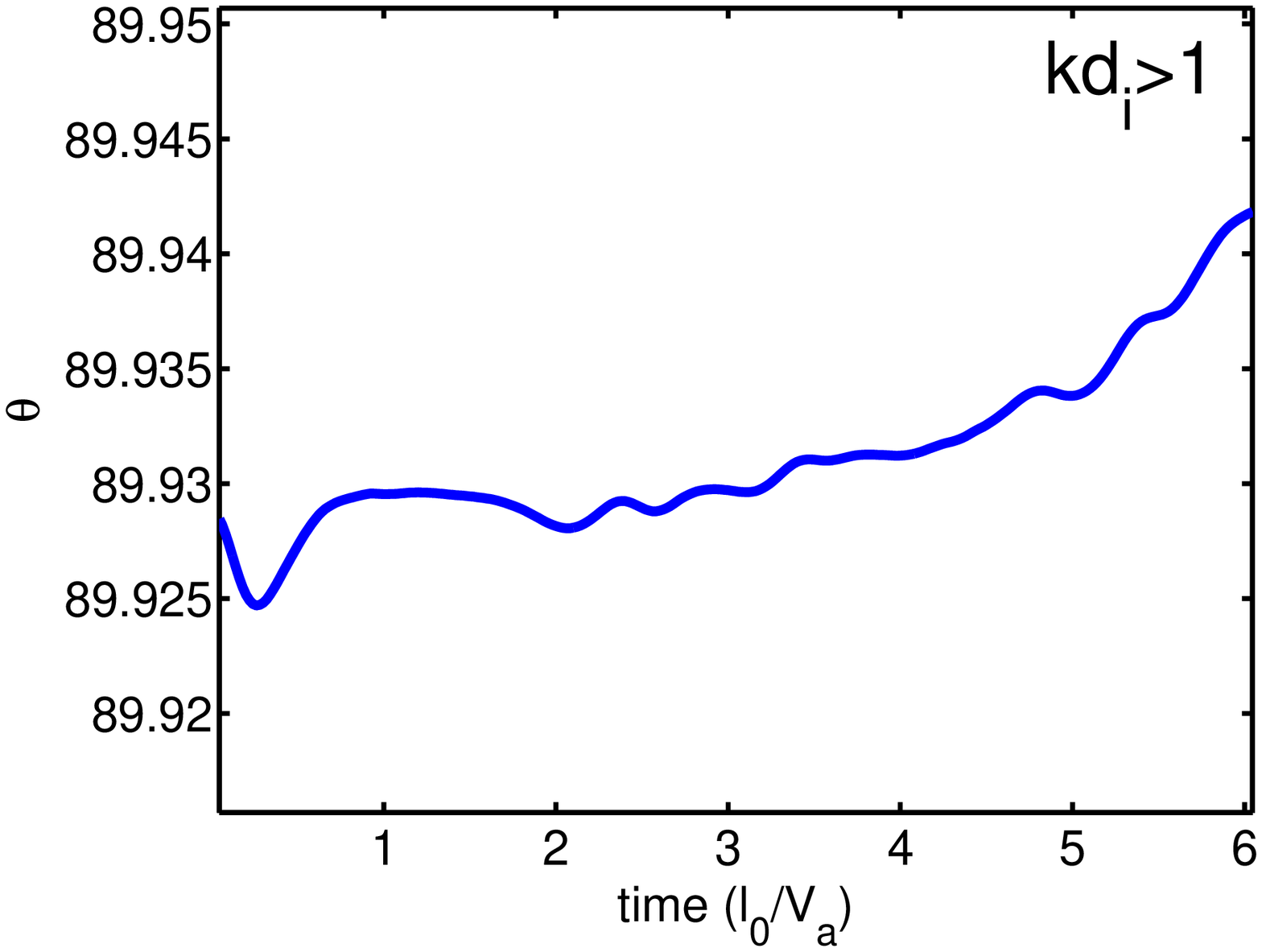}
  \caption{ A progressive decrease in the angle of alignment from
  $90^\circ$ indicates the eventual weakening of ${\bf V} \times {\bf
  B}$ nonlinear interactions.  Evolution of the degree of alignment in
  the $kd_i>1$ regime of Hall MHD. Small scale fluctuations possess
  orthogonal velocity and magnetic fields.}
\end{figure}

\section{3. anisotropic cascades}
To measure the degree of anisotropic cascades (or spectral anisotropy)
by employ the following diagnostics to monitor the evolution of the
$k_\perp$ mode in time. The averaged $k_\perp$ mode is determined by
averaging over the entire turbulent spectrum weighted by $k_\perp$,
thus $ \langle k_\perp(t)\rangle = ({\sum_k |k_\perp
Q(k,t)|^2}/{\sum_k |Q(k,t)|^2})^{1/2}.$ Here $\langle \cdots \rangle$
represents an average over the entire Fourier spectrum, $
k_\perp=({k_x^2+k_y^2})^{1/2}$ and $Q$ represents any of $B, V, \rho$,
$\nabla \times {\bf B}$ and $\nabla \times {\bf V}$.  Similarly, the
evolution of the $k_\parallel$ mode is determined by the relation,
$\langle k_\parallel(t) \rangle= ({\sum_k |k_\parallel
Q(k,t)|^2}/{\sum_k |Q(k,t)|^2})^{1/2}.$ It is clear from these expressions
that the $\langle k_\perp(t)\rangle$ and $\langle
k_\parallel(t) \rangle$ modes exhibit isotropy when $\langle
k_\perp(t)\rangle \simeq \langle k_\parallel(t)
\rangle$. Any deviation from this equality corresponds to spectral
anisotropy. We follow the evolution of the $\langle k_\perp(t)\rangle$
and $\langle k_\parallel(t) \rangle$ modes in our simulations.  Our
simulation results describing the evolution of $\langle k_\perp
\rangle$ and $\langle k_\parallel(t) \rangle$ modes are shown in
\Fig{aniso}. It is evident from \Fig{aniso} that the initially
isotropic modes $\langle k_\perp(t)\rangle \simeq \langle
k_\parallel(t) \rangle$ gradually evolve towards a highly anisotropic
state in that spectral transfer preferentially occurs in the $\langle
k_\perp(t)\rangle$ mode, and is suppressed in $\langle k_\parallel(t)
\rangle$ mode. Consequently, spectral transfer in the $\langle
k_\perp(t)\rangle$ mode dominates the nonlinear evolution of
fluctuations in Hall MHD, and mode structures become elongated along
the mean magnetic field or $z$-direction. Hence nonlinear interactions
led by the nonlinear terms in the presence of background gradients
lead to anisotropic turbulent cascades in the inertial range turbulent
spectra.

 \Fig{aniso} illustrates anisotropy corresponding to an
averaged $k$ mode but the anisotropy exhibited by the small and large
scale ${\bf B}$ and ${\bf v}$ fluctuations is not distinctively clear,
nor is the degree of anisotropy in ${\bf B}$ and ${\bf v}$ fields
clear from \Fig{aniso}.  The scale dependece of turbulent anisotropy
is described in \Fig{aniso2}. \Fig{aniso2} shows discrepancy in
$k_\perp$ and $k_\parallel$ is prominent at the smaller $k$'s.  This
essentially means that the large scale turbulent fluctuations are more
anisotropic than the smaller ones in a regime where characteristic
length scales are smaller than $d_i$ i.e. $kd_i > 1$.  It further
appears from \Fig{aniso2} that the smaller scales in the $kd_i > 1$
are virtually {\em unaffected} by anisotropic kinetic Alfv\'en waves
that propagate along the externally imposed mean magnetic field $B_0$.
Turbulent fluctuations with small characteristic scales in the $kd_i >
1$ regime of Hall MHD are not affected by the mean magnetic field or
kinetic Alfv\'en waves. This leads us to conjecture that small scale
turbulence in the $kd_i > 1$ regime behaves essentially
hydrodynamically i.e.  as eddies independent of the mean magnetic
field or collisionless magnetized waves.  Thus large and smaller
turbulent length scales evolve differently in the $kd_i > 1$ regime of
Hall MHD turbulence.

\section{3. Dynamical alignment  of  fluctuations}
To understand the strength of the nonlinear interactions in Hall MHD
solar wind plasma, we determine the degree of alignment of the
velocity and magnetic field fluctuations by defining the following
alignment parameter \cite{podesta} that spans the entire $k$-spectrum
in both the $kd_i>1$ (Hall MHD) and $kd_i<1$ (usual MHD) regimes.
$ \Theta(t) = \cos^{-1} \left({\sum_{\bf k} {\bf V}_{\bf k}
(t) \cdot {\bf B}_{\bf k} (t)}/{\sum_{\bf k}|{\bf V}_{\bf k}(t)| |{\bf
B}_{\bf k}(t)|} \right).$ The summation is determined from the modes
by summing over the entire spectrum. In this sense, the alignment
parameter depicts an average evolution of the alignment of velocity
relative to the magnetic field fluctuations. Note carefully that this
alignment can vary locally from smaller to larger scales, but the
averaging (i.e. summing over the entire spectrum) rules out any such
possibility in our simulations. Nonetheless, $\Theta$ defined as above
enables us to quantitatively measure the average alignment of the
magnetic and velocity field fluctuations while the nonlinear
interactions evolve in a turbulent solar wind plasma.

 The Alfv\'enic cascade regime of MHD turbulence $kd_i<1$ in the solar
wind plasma possesses relatively large scales ($kd_i<1$) in which the
velocity and magnetic field fluctuations are observed to be somewhat
obliquely aligned. Hence our simulations show that the angle of
alignment evolves towards $\Theta < 90^\circ$, as depicted
in \Fig{angle}.  Hence the strength of the nonlinear interactions
corresponding to the ${\bf V} \times {\bf B}$ nonlinearity is
relatively weak. This result is to be contrasted with characteristic
turbulent length scales in the $kd_i>1$ regime. The angle of alignment
for the smaller scales corresponding to the $kd_i>1$ regime is shown
in \Fig{angle}. Significant differences are apparent in the angle of
alignments associated with the large and small scales \Fig{angle}.  It
appears from our simulations that the small scale fluctuations
($kd_i>1$) are nearly orthogonal as seen in \Fig{angle}. By contrast,
the large scale fluctuations ($kd_i<1$) in \Fig{angle} show a
significant departure from the orthogonality.

\section{4. Conclusions}
In summary, we have investigated the nonlinear and turbulent behavior
of a two fluid, compressible, three dimensional Hall MHD model.  In
the presence of a large scale mean background magnetic field, small
scale turbulent fluctuations exhibit anisotropic power spectra close
to $k^{-7/3}$.  We find that the long length scales in the $kd_i>1$
KAW regime are more anisotropic compared to the shorter scales.
Dynamical alignment and angular distribution of turbulence velocity
and magnetic field fluctuations is found to play a critical role in
determining the degree of nonlinear interactions.  We find that
characteristic turbulent flucutations in the $kd_i>1$ regime relax
towards orthogonality, so that most of turbulent scale fluctuations
have velocity and magnetic fields that are nearly orthogonal ,
i.e. ${\bf V} \perp {\bf B}$.  For large scale fluctuations
corresponding to the MHD regime, magnetic and velocity fields are not
perfectly orthogonal, being instead on average nearly $70^\circ$ to
each other. By contrast, small scale fluctuations in the $kd_i>1$ KAW
regime exhibit nearly perfect orthogonality in that the average
magnetic and velocity fields make an angle of nearly $90^\circ$ with
respect to each other.

The spectral properties of nonlinear Hall MHD are particularly
relevant for understanding the observed solar wind and heliospheric
turbulence.  Hall MHD may also be useful for understanding multi-scale
electromagnetic fluctuations and magnetic field reconnection in the
Earth's magnetosphere and in laboratory plasmas.


The support of NASA(NNG-05GH38) and NSF (ATM-0317509) grants is  acknowledged.





\IfFileExists{\jobname.bbl}{}
 {\typeout{}
  \typeout{******************************************}
  \typeout{** Please run "bibtex \jobname" to optain}
  \typeout{** the bibliography and then re-run LaTeX}
  \typeout{** twice to fix the references!}
  \typeout{******************************************}
  \typeout{}
 }


\end{document}